\documentclass[onecolumn,showpacs,preprintnumbers,amsmath,amssymb,aps,pra,groupedaddress]{revtex4-2}
\usepackage{multirow,diagbox}  
\usepackage{siunitx}
\usepackage[colorlinks,citecolor=blue,linkcolor=red,hyperindex,CJKbookmarks]{hyperref}
\usepackage{amsmath}
\usepackage{graphicx}
\usepackage{dcolumn}
\usepackage{subfigure}
\usepackage{mathrsfs}
\usepackage{amssymb}
\usepackage{amsfonts}
\usepackage{makecell}
\usepackage{color}
\begin{document}
	\title{Revisiting the origin of neutrino flavor transformations}
	\author{Shi-Biao Zheng}\thanks{%
		E-mail: t96034@fzu.edu.cn}
	\address{Department of Physics, Fuzhou University, Fuzhou 350108, China}
	\date{\today}

\begin{abstract}
To account for neutrino oscillations, it was postulated that the neutrino
has nonvanishing mass and each flavor eigenstate is formed by a quantum
superposition of three distinct mass eigenstates, whose probability
amplitudes interfere with each other during its propagation. However, I find
that a neutrino or antineutrino produced by the decay of an unstable
particle cannot be in such a superposition, as different mass eigenstates,
if they exist, are necessarily correlated with different momentum states of
the composite system produced by the decay, which would destroy the quantum
coherence among these mass eigenstates. I further find that the states of a
neutrino and an electron become nonseparable after their charged-current
interaction. This nonseparability leads to decoherence for neutrinos
propagating in matter, but was not taken into consideration in previous
investigations of the matter effect. Due to this decoherence, the deficit of
solar electron neutrinos cannot exceed 1/2 based on the aforementioned
postulation even if the corresponding superposition of mass eigenstates can
be produced. These results unambiguously show that the origin of neutrino
flavor transformations needs to be revisited. I propose an alternative
mechanism that can reasonably account for neutrino transformations. It is
based on virtual excitation of the Z bosonic field diffusing over the space.
During the propagation, the neutrino can continually excite and then
immediately re-absorb a virtual Z boson. This virtual bosonic excitation
produces a backaction on the neutrino, enabling it to oscillate among three
flavors. When the neutrino propagates in matter, its behavior is determined
by the competition between the coherent flavor transformation and
decoherence effect resulting from scatterings.
\end{abstract}

\vskip0.5cm

\narrowtext
\maketitle
\section{INTRODUCTION}

The discovery of neutrino flavor transformations represents one of the most
striking advancements of physical science achieved during the past few
decades [1-27]. Such phenomena are commonly believed to show that neutrinos
have nonvanishing mass, which is beyond the standard model [28,29]. To
account for these phenomena, it is assumed that there exist three distinct
neutrino mass eigenstates. Each flavor eigenstate corresponds to a
quantum-mechanical superposition of these mass eigenstates, and vice versa
[30]. During the propagation of the neutrino, the probability amplitude
associated with each mass eigenstate accumulates a phase that is
approximately proportional to the square of the corresponding mass. The
time-evolving phase differences among these probability amplitudes give
arise to flavor oscillations, as a consequence of their quantum interference.

This interpretation is valid only when the neutrino can be in the
superposition state at the production. Previously, it was realized that the
neutrino emitted by an unstable particle with a definite momentum is
necessarily entangled with the particles accompanying the neutrino at the
production [31-33]. This entanglement would destroy the coherence between
the neutrino's mass eigenstates. To overcome this problem, it was argued
that the neutrino can be disentangled with the accompanying particles when
the momentum uncertainty of the unstable particle is sufficiently large
[31-33]. However, I find that different mass eigenstates of such a neutrino,
if they exist, are necessarily correlated with different joint momentum
states of the entire system produced by the decay, including the neutrino
and the accompanying particles, which prohibits occurrence of interference
effects between the mass eigenstates. I further find that the previous
interpretation of the matter effect has overlooked the critical fact that
the states of the neutrino and electron become nonseparable after their
charged-current (CC) interaction [34-36]. Due to this nonseparability, the
solar $^{8}$B neutrino cannot adiabatically evolve from a quantum
superposition of different mass eigenstates to a pure mass eigenstate, which
was assumed to be responsible for the observed 2/3 deficit. I finally
propose an alternative mechanism, based on the virtual excitation of the Z
bosonic field, to interpret neutrino flavor transformations observed in
different experiments.

\section{NO-GO THEOREM REGARDING PRODUCTION OF SUPERPOSITIONS OF MASS
EIGENSTATES}

This section is devoted to the proof of the no-go theorem that the neutrino
or antineutrino emitted by an unstable particle cannot be in a
quantum-mechanical superposition of different mass eigenstates. For the sake
of clarity, I will illustrate this theorem with the $\beta $ decay of a
neutron, but the conclusion holds for other weak CC decays.

{\sl lemma 1}: For the $\beta $ decay, when there is no mass-momentum
entanglement, the produced electron antineutrino has a definite mass.

The wavefunction of the neutron, before undergoing the $\beta $ decay, can
be expanded as%
\begin{equation}
\left\vert \psi _{n}\right\rangle =\int \varphi ({\bf P}_{n})d^{3}{\bf P}%
_{n}\left\vert {\bf P}_{n}\right\rangle ,
\end{equation}%
where $\left\vert {\bf P}_{n}\right\rangle $ denotes the momentum eigenstate
of the neutron with the eigenvalue ${\bf P}_{n}$. Suppose that the electron
antineutrino produced by the $\beta $ decay possesses three different mass
eigenstates, which are not entangled with different momentum eigenstates of
the antineutrino, proton, and electron. Then the state of the composite
system can be written as%
\begin{equation}
\left\vert \psi _{\nu +p+e}\right\rangle =\int d^{3}{\bf P}_{\nu }d^{3}{\bf P%
}_{p}d^{3}{\bf P}_{e}F({\bf P}_{\nu },{\bf P}_{p},{\bf P}_{e})\left\vert 
{\bf P}_{\nu },{\bf P}_{p},{\bf P}_{e}\right\rangle (%
\mathop{\displaystyle\sum}%
\limits_{j=1}^{3}C_{j}\left\vert \stackrel{-}{\nu }_{j}\right\rangle ),
\end{equation}%
where $\left\vert \stackrel{-}{\nu }_{j}\right\rangle $ denotes the $j$th
eigenstate for the electron antineutrino. $\left\vert {\bf P}_{\nu },{\bf P}%
_{p},{\bf P}_{e}\right\rangle $ represents joint momentum eigenstate of the
entire system, where the antineutrino, proton, and electron are all in their
momentum eigenstates with the eigenvalues ${\bf P}_{\nu }$, ${\bf P}_{p}$,
and ${\bf P}_{e}$ respectively. The joint probability amplitude distribution 
$F({\bf P}_{\nu },{\bf P}_{p},{\bf P}_{e})$ satisfies the normalization
condition 
\begin{equation}
\int d^{3}{\bf P}_{\nu }d^{3}{\bf P}_{p}d^{3}{\bf P}_{e}\left\vert F({\bf P}%
_{\nu },{\bf P}_{p},{\bf P}_{e})\right\vert ^{2}=1.
\end{equation}%
The joint momentum of the entire antineutrino-proton-electron system is
essentially in a superposition of infinitely many components, which implies
that both the total momentum and energy are undeterministic. Despite these
uncertainties, the energy and momentum conservation laws are still satisfied
for each momentum component of the wave function, as correctly pointed out
in Ref. [31].

We here consider a specific component, denoted as $\left\vert {\bf P}_{\nu
}^{0},{\bf P}_{p}^{0},{\bf P}_{e}^{0}\right\rangle $. The momentum
conservation law implies that this momentum component originates from the
neutron momentum component $\left\vert {\bf P}_{n}^{0}\right\rangle $, with 
\begin{equation}
{\bf P}_{n}^{0}={\bf P}_{\nu }^{0}+{\bf P}_{p}^{0}+{\bf P}_{e}^{0}.
\end{equation}%
The energies of the neutron, proton, and electron associated with the
component $\left\vert {\bf P}_{\nu }^{0},{\bf P}_{p}^{0},{\bf P}%
_{e}^{0}\right\rangle $ are given by 
\begin{eqnarray}
E_{n}^{0} &=&\sqrt{m_{n}^{2}+(P_{n}^{0})^{2}},  \nonumber \\
E_{p}^{0} &=&\sqrt{m_{p}^{2}+(P_{p}^{0})^{2}},  \nonumber \\
E_{e}^{0} &=&\sqrt{m_{e}^{2}+(P_{e}^{0})^{2}},
\end{eqnarray}%
where $m_{n}$, $m_{p}$, and $m_{e}$ are the masses of the neutron, proton,
and electron, respectively. According to the energy conservation law, the
antineutrino's energy associated with the component $\left\vert {\bf P}_{\nu
}^{0},{\bf P}_{p}^{0},{\bf P}_{e}^{0}\right\rangle $ is definite, given by $%
E_{\nu }^{0}=E_{n}^{0}-E_{p}^{0}-E_{e}^{0}$. Consequently, the
antineutrino's mass is also definite, which is equal to $m_{\nu }=\sqrt{%
(E_{\nu }^{0})^{2}-(P_{\nu }^{0})^{2}}$. This leads to $m_{j}=m_{\nu }$ when 
$C_{j}\neq 0$. Such a conclusion is inconsistent with the postulation that
each flavor eigenstate is a linear superposition of three different mass
eigenstates. If the flavor oscillations are caused by nonzero mass
differences, this state cannot exhibit any oscillatory behavior.

{\sl lemma 2}: For the $\beta $ decay, different mass eigenstates of the
produced electron antineutrino are necessarily correlated with different
joint antineutrino-proton-electron momentum states that are orthogonal to
each other.

Generally, the wavefunction of the entire system produced by the $\beta $
decay can be written in the form of

\begin{equation}
\left\vert \psi \right\rangle =%
\mathop{\displaystyle\sum}%
\limits_{j}\int_{{\bf \sigma }_{j}}d^{3}{\bf P}_{\nu ,j}d^{3}{\bf P}%
_{p,j}d^{3}{\bf P}_{e,j}G({\bf P}_{\nu ,j},{\bf P}_{p,j},{\bf P}%
_{e,j})\left\vert {\bf P}_{\nu ,j},{\bf P}_{p,j},{\bf P}_{e,j}\right\rangle
\left\vert \stackrel{-}{\nu }_{j}\right\rangle ,
\end{equation}%
where ${\bf \sigma }_{j}$ denotes the distribution region of the joint
antineutrino-proton-electron momentum associated with $\left\vert \stackrel{-%
}{\nu }_{j}\right\rangle $. In order to satisfy the condition $m_{1}\neq
m_{2}\neq m_{3}$, there should not be any overlapping between the momentum
distribution regions associated with different mass eigenstates, i.e., ${\bf %
\sigma }_{j}\cap $ ${\bf \sigma }_{k}=\oslash $ for $j\neq k$. This can be
interpreted as follows. Suppose that there is an overlapping between the
regions ${\bf \sigma }_{j}$ and ${\bf \sigma }_{k}$ with $j\neq k$. Then,
according to the aforementioned analysis, both $m_{j}$ and $m_{k}$ can be
uniquely determined by a specific joint momentum component $\left\vert {\bf P%
}_{\nu }^{0},{\bf P}_{p}^{0},{\bf P}_{e}^{0}\right\rangle $ that falls
within the overlapping regime. This implies $m_{j}=m_{k}$ when ${\bf \sigma }%
_{j}\cap $ ${\bf \sigma }_{k}\neq \oslash $.

For such an entangled state, when the momentum states are traced out, the
mass degree of freedom is left in a classical mixture, described by the
density operator 
\begin{eqnarray}
\rho _{\nu } &=&Tr_{{\bf P}_{\nu },{\bf P}_{n},{\bf P}_{e}}\left\vert \psi
\right\rangle \left\langle \psi \right\vert  \nonumber \\
&=&\int d^{3}{\bf P}_{\nu }d^{3}{\bf P}_{p}d^{3}{\bf P}_{e}\left\langle {\bf %
P}_{\nu },{\bf P}_{p},{\bf P}_{e}\right\vert \left. \psi \right\rangle
\left\langle \psi \right\vert \left. {\bf P}_{\nu },{\bf P}_{p},{\bf P}%
_{e}\right\rangle  \nonumber \\
&=&%
\mathop{\displaystyle\sum}%
\limits_{j,k}D_{j,k}\left\vert \stackrel{-}{\nu }_{j}\right\rangle
\left\langle \stackrel{-}{\nu }_{k}\right\vert ,
\end{eqnarray}%
with 
\begin{eqnarray}
D_{j,k} &=&\int_{{\bf \sigma }_{j}}d^{3}{\bf P}_{\nu ,j}d^{3}{\bf P}%
_{p,j}d^{3}{\bf P}_{e,j}\int_{{\bf \sigma }_{k}}d^{3}{\bf P}_{\nu ,k}d^{3}%
{\bf P}_{p,k}d^{3}{\bf P}_{e,k}  \nonumber \\
&&G({\bf P}_{\nu ,j},{\bf P}_{p,j},{\bf P}_{e,j})G^{\ast }({\bf P}_{\nu ,k},%
{\bf P}_{p,k},{\bf P}_{e,k})  \nonumber \\
&&\left\langle {\bf P}_{\nu ,k},{\bf P}_{p,k},{\bf P}_{e,k}\right\vert
\left. {\bf P}_{\nu ,j},{\bf P}_{p,j},{\bf P}_{e,j}\right\rangle .
\end{eqnarray}%
Since ${\bf \sigma }_{j}\cap $ ${\bf \sigma }_{k}=\oslash $ for $j\neq k$,
each of the joint momentum eigenstates in the region ${\bf \sigma }_{j}$ is
orthogonal to all the momentum eigenstates in ${\bf \sigma }_{k}$. This
implies $\left\langle {\bf P}_{\nu ,k},{\bf P}_{p,k},{\bf P}%
_{e,k}\right\vert \left. {\bf P}_{\nu ,j},{\bf P}_{p,j},{\bf P}%
_{e,j}\right\rangle =0$ throughout the integral region ${\bf \sigma }%
_{j}\otimes {\bf \sigma }_{k}$ for $j\neq k$. Therefore, we have%
\begin{equation}
\rho _{\nu }=%
\mathop{\displaystyle\sum}%
\limits_{j}D_{j,j}\left\vert \stackrel{-}{\nu }_{j}\right\rangle
\left\langle \stackrel{-}{\nu }_{j}\right\vert ,
\end{equation}%
where%
\begin{equation}
D_{j,j}=\int_{{\bf \sigma }_{j}}d^{3}{\bf P}_{\nu ,j}d^{3}{\bf P}_{p,j}d^{3}%
{\bf P}_{e,j}\left\vert G({\bf P}_{\nu ,j},{\bf P}_{p,j},{\bf P}%
_{e,j})\right\vert ^{2}.
\end{equation}%
In other words, the quantum coherence among the mass eigenstates is
destroyed by their quantum entanglement with different joint momentum states.

The aforementioned proof indicates that when the antineutrino is not
entangled with its accompanying particles, its mass eigenstates, if they
exist, would be necessarily correlated with different momenta of the
antineutrino itself. This mass-momentum correlation would destroy the
quantum coherence between the neutrino's mass eigenstates, which was
overlooked by other authors in previous investigations of entanglement and
coherence associated with neutrino oscillations [31-33]. This decoherece
effect can be further interpreted in terms of complementarity [37-45]. The
information about the mass eigenstate of the antineutrino is encoded in its
momentum. Consequently, one can determine which eigenstate the antineutrino
is\ in by measuring its momentum in principle, which is sufficient to
destroy the coherence among the mass eigenstates. Such a decoherence effect
does not depend upon whether or not the momentum is actually measured.

This decoherece effect can also be illustrated in the position
representation. When the antineutrino is not entangled with other particles,
its wavefunction in this representation evolves as%
\begin{equation}
\left\vert \varphi _{\nu }(t)\right\rangle =%
\mathop{\displaystyle\sum}%
\limits_{j}\int d^{3}{\bf r}f_{j}({\bf r},t)\left\vert {\bf r}\right\rangle
\left\vert \stackrel{-}{\nu }_{j}\right\rangle ,
\end{equation}%
where 
\begin{equation}
f_{j}({\bf r},t)=(2\pi )^{-3/2}\int_{{\bf \sigma }_{j}}d^{3}{\bf P}_{\nu
,j}G({\bf P}_{\nu ,j})e^{i({\bf P}_{\nu ,j}\cdot {\bf r-}E_{j}t)},
\end{equation}%
$E_{j}=\sqrt{p_{j}^{2}+m_{j}^{2}}$ with $m_{j}$ being the mass of the $j$th
mass eigenstate, and $\left\vert {\bf r}\right\rangle $ denotes the position
eigenstate. The coherence between $\left\vert \stackrel{-}{\nu }%
_{j}\right\rangle $ and $\left\vert \stackrel{-}{\nu }_{k}\right\rangle $
manifested on the detection of the antineutrino is given by 
\begin{eqnarray}
C_{j,k} &=&\int_{D}d^{3}{\bf r}f_{j}({\bf r},t)f_{k}^{\ast }({\bf r},t) 
\nonumber \\
&=&(2\pi )^{-3}\int_{{\bf \sigma }_{j}}d^{3}{\bf P}_{\nu ,j}\int_{{\bf %
\sigma }_{k}}d^{3}{\bf P}_{\nu ,k}G({\bf P}_{\nu ,j})G^{\ast }({\bf P}_{\nu
,k})\int_{D}d^{3}{\bf r}e^{i[({\bf P}_{\nu ,j}-{\bf P}_{\nu ,k})\cdot {\bf %
r-(}E_{k}-E_{j})t]},
\end{eqnarray}%
where $D$ is the detection region of the antineutrino. When the size of the
detector is much larger than that of the antineutrino's wavepacket, $%
\int_{D}d^{3}{\bf r}e^{i({\bf P}_{\nu ,j}-{\bf P}_{\nu ,k})\cdot {\bf r}}$
can be well approximated by taking the integral over the whole space. As $%
{\bf P}_{\nu ,j}\neq {\bf P}_{\nu ,k}$, such an integral is zero, which
implies that the coherence $C_{j,k}$ vanishes. This result can also be
understood in terms of the position-dependent phase difference between $%
\left\vert \stackrel{-}{\nu }_{j}\right\rangle $ and $\left\vert \stackrel{-}%
{\nu }_{k}\right\rangle $ owing to the associated momentum difference. We
note that such phase differences were also included in previous
investigations, exemplified by the statement "{\sl Since }$p_{x,i}\neq
p_{x,j}${\sl , phase differences exist between the components at the point
of detection}" in Ref. [32]. However, the detection of neutrinos or
antineutrinos cannot be restricted to a single point. The interference
effects of mass eigenstates (internal degree of freedom) would be averaged
out when integrating the position (external degree of freedom) over the
large volume of the detector, as the position-averaged value of the phase
factor caused by the corresponding momentum difference is zero.

\section{NEUTRINO-ELECTRON MIXTURE PRODUCED BY THE CHARGED-CURRENT
INTERACTION}

The no-go theorem regarding production of superpositions of mass eigenstates
is applicable to neutrinos or antineutrinos produced in the weak
charged-current decay of other unstable particles, including mesons and
muons. It also holds solar $^{8}$B neutrinos, which are produced by the
reaction [2] 
\begin{equation}
^{8}\text{B}\rightarrow ^{8}\text{Be}+e^{+}+\nu _{e}.
\end{equation}%
I further note that even if solar $^{8}$B neutrinos can be initially in a
superposition of mass eigenstates, they cannot adiabatically evolve into a
pure mass eigenstate, where the population of the electron flavor eigenstate
was assumed to be about 1/3 [2], as will be detailedly interpreted below.

Previously, the flavor transformation of solar $^{8}$B neutrinos was
attributed to the matter effect proposed by Mikheyev and Smirnov by
extending the idea of Wolfenstein, referred to as the MSW effect [34-36].
This effect originates from the CC interaction between the electron neutrino
and the background electrons in matter, which can be described by the
effective Hamiltonian, 
\begin{equation}
H_{cc}=\frac{G_{F}}{\sqrt{2}}{\bf \nu }_{e}^{+}\gamma _{4}\gamma _{\lambda
}(1+\gamma _{5}){\bf ee}^{\dagger }\gamma _{4}\gamma _{\lambda }(1+\gamma
_{5}){\bf \nu }_{e},
\end{equation}%
where ${\bf \nu }_{e}$ and ${\bf e}$ denote the fields associated with the
electron neutrino and electron, respectively. Using the Fierz
transformation, the Hamiltonian was rewritten in the form of%
\begin{equation}
H_{cc}^{\prime }=\frac{G_{F}}{\sqrt{2}}{\bf \nu }_{e}^{+}\gamma _{4}\gamma
_{\lambda }(1+\gamma _{5}){\bf \nu }_{e}{\bf e}^{\dagger }\gamma _{4}\gamma
_{\lambda }(1+\gamma _{5}){\bf e}.
\end{equation}%
Then the electron field was considered as a static background, whose state
is not affected by the CC interaction, so that ${\bf e}^{\dagger }\gamma
_{4}\gamma _{\lambda }(1+\gamma _{5}){\bf e}$ can be replaced with $\delta
_{\lambda ,4}N_{e}$, where $N_{e}$ is the number density of electrons. With
this treatment, the Hamiltonian $H_{cc}^{\prime }$ is effectively equivalent
to an external potential for the neutrino, given by $V=\sqrt{2}N_{e}G_{F}$.

To derive the MSW effect, it was further supposed that each neutrino flavor
eigenstate is formed by a linear superposition of three mass eigenstates 
\begin{equation}
\left\vert \nu _{\alpha }\right\rangle =%
\mathop{\displaystyle\sum}%
\limits_{j=1}^{3}U_{\alpha j}\left\vert \nu _{j}\right\rangle ,
\end{equation}%
where $j$ labels the mass eigenstate, and $\alpha =e,\mu ,\tau $ denotes the
flavor of the neutrino. $\left\vert \nu _{e}\right\rangle $ was assumed to
be approximated by a superposition of $\left\vert \nu _{1}\right\rangle $
and $\left\vert \nu _{2}\right\rangle $, i.e., $U_{e3}\simeq 0$ [2]. Under
this assumption, neither $H_{cc}^{\prime }$ nor the free Hamiltonian can
couple $\left\vert \nu _{e}\right\rangle $ to $\left\vert \nu
_{3}\right\rangle $, and thus the population of $\left\vert \nu
_{3}\right\rangle $ can be neglected for the initial state $\left\vert \nu
_{e}\right\rangle $. Then the dynamics can be described in a two-dimensional
subspace $\{\left\vert \nu _{e}\right\rangle ,\left\vert \nu _{\beta
}\right\rangle \}$, where%
\begin{equation}
\left\vert \nu _{\beta }\right\rangle ={\cal N}_{\beta }(U_{\tau
3}\left\vert \nu _{\mu }\right\rangle -U_{\mu 3}\left\vert \nu _{\tau
}\right\rangle ),
\end{equation}%
with ${\cal N}_{\beta }=\left( \left\vert U_{\tau 3}\right\vert
^{2}+\left\vert U_{\mu 3}\right\vert ^{2}\right) ^{-1/2}$. Within this
subspace, the Hamiltonian can be approximately expressed as%
\begin{equation}
H\simeq V\left\vert \nu _{e}\right\rangle \left\langle \nu _{e}\right\vert +%
\mathop{\displaystyle\sum}%
\limits_{\eta ,\xi =e,\beta }M_{\eta ,\xi }(p)\left\vert \nu _{\eta
}\right\rangle \left\langle \nu _{\xi }\right\vert ,
\end{equation}%
where%
\begin{equation}
M_{\eta ,\xi }(p)\simeq 
\mathop{\displaystyle\sum}%
\limits_{j=1,2}\frac{m_{j}^{2}}{2p}U_{\eta j}^{\ast }U_{\xi j},
\end{equation}%
with $U_{\beta j}={\cal N}_{\beta }(U_{\tau 3}U_{\mu j}-U_{\mu 3}U_{\tau j})$%
. Here $p$ denotes the neutrino momentum, which is much larger than the mass
($m_{j}$) associated with each mass eigenstate. The trivial common energy,
described by $p%
\mathop{\displaystyle\sum}%
\limits_{j=1,2}\left\vert \nu _{j}\right\rangle \left\langle \nu
_{j}\right\vert $, has been discarded. When $V\gg m_{j}^{2}/2p$, the
electron flavor approximately coincides with the eigenstate of the
Hamiltonian with the larger eigenenergy. If the electron number density is
changed sufficiently slowly, the neutrino adiabatically follows the
corresponding Hamiltonian eigenstate during its propagation. On the solar
surface, $V$ can be neglected as compared to $m_{j}^{2}/2p$ so that the
eigenstates of the Hamiltonian coincide with the mass eigenstates. This
implies that the initial electron flavor eigenstate evolves to the mass
eigenstate with the larger mass ($\left\vert \nu _{2}\right\rangle $) when
the neutrino reaches the solar surface. This mass eigenstate remains
invariant until being detected on the Earth. The resulting probability $%
P_{\left\vert \nu _{e}\right\rangle \rightarrow \left\vert \nu
_{e}\right\rangle }$ is approximately equal to $\left\vert U_{2e}^{\dagger
}\right\vert ^{2}$, which was assumed to be about $1/3$ [2].

This treatment has overlooked the crucial fact that the CC reaction leads to
neutrino-electron entanglement when the neutrino is in a superposition of
the electron flavor eigenstate and the other two flavor eigenstates before
the reaction. As the neutrino and the electron can be transformed into each
other by their CC reaction, it helps to make the presentation more clear to
refer the original neutrino and the original electron to as particle 1 and
particle 2, respectively. The CC reaction transforms the state $\left\vert
\nu _{e},{\bf p}_{1}\right\rangle _{1}\left\vert e,{\bf p}_{2}\right\rangle
_{2}$ into $\left\vert e,{\bf p}_{3}\right\rangle _{1}\left\vert \nu _{e},%
{\bf p}_{4}\right\rangle _{2}$, where the subscripts "1" and "2" outside the
kets label the two particles, and ${\bf p}_{1}$ (${\bf p}_{3}$) and ${\bf p}%
_{2}$ (${\bf p}_{4}$) are their momenta before (after) the reaction. If
particle 1 is initially in the flavor eigenstate $\left\vert \nu
_{e}\right\rangle $ and ${\bf p}_{1}={\bf p}_{4}$, the Fierz rearranging is
equivalent to relabeling the two particles, which does not cause any
problem. However, when it is initially in a superposition of $\left\vert \nu
_{e}\right\rangle $ and $\left\vert \nu _{\beta }\right\rangle $, it will be
entangled with particle 2 by the CC reaction. To illustrate this point, we
suppose that the two-particle system is initially in the state 
\begin{equation}
\left\vert \psi _{0}\right\rangle =(C_{e}\left\vert \nu _{e},{\bf p}%
_{1}\right\rangle _{1}+C_{\beta }\left\vert \nu _{\beta },{\bf p}%
_{1}\right\rangle _{1})\left\vert e,{\bf p}_{2}\right\rangle _{2}.
\end{equation}%
In this case, the CC reaction actually corresponds to a conditional
dynamics, by which particle 1 exchanges its state with particle 2 when it is
initially in the electron flavor eigenstate, but nothing occurs if it is
initially in the other two flavor eigenstates. This conditional state
swapping evolves the system to the entangled state 
\begin{equation}
\left\vert \psi \right\rangle =C_{e}\left\vert e,{\bf p}_{3}\right\rangle
_{1}\left\vert \nu _{e},{\bf p}_{4}\right\rangle _{2}+C_{\beta }\left\vert
\nu _{\beta },{\bf p}_{1}\right\rangle _{1}\left\vert e,{\bf p}%
_{2}\right\rangle _{2}.
\end{equation}%
It should be noted that the electron transformed from the neutrino does not
have the same momentum as the original electron, i.e., ${\bf p}_{3}\neq {\bf %
p}_{2}$. Such momentum differences have been used to identify
neutrino-electron scattering events in SNO experiments [2]. This quantum
entanglement is masked by the Fierz rearranging and the subsequent
replacement of the electron part in the Hamiltonian with a number. We
further note that the Fierz rearranging is valid for calculation of the $e$-$%
\nu _{e}$ scattering amplitude, which is irrelevant to the quantum coherence
between $\left\vert \nu _{e}\right\rangle $ and $\left\vert \nu _{\beta
}\right\rangle $. However, it overlooks the fact that $\left\vert \nu
_{e}\right\rangle $ and $\left\vert \nu _{\beta }\right\rangle $ are carried
by different particles after the CC reaction, which is essential for correct
description of the neutrino state evolution in matter. In other words, the
states of the two particles are no longer separable after their CC
interaction, so that the electrons participating in such interactions cannot
be treated as a static background for the neutrino, and their effects cannot
be modeled as a potential for the propagating neutrino.

Due to the quantum entanglement, each of the two particles is essentially in
a mixture of the neutrino and electron states. This critical point can be
illustrated more clearly by the reduced density operators for these
particles, each obtained by tracing out the degree of freedom of the other
particle, given by%
\begin{eqnarray}
\rho _{1} &=&Tr_{2}(\left\vert \psi \right\rangle \left\langle \psi
\right\vert )  \nonumber \\
&=&\left\vert C_{e}\right\vert ^{2}\left\vert e,{\bf p}_{3}\right\rangle
_{1}\left\langle e,{\bf p}_{3}\right\vert +\left\vert C_{\beta }\right\vert
^{2}\left\vert \nu _{\beta },{\bf p}_{1}\right\rangle _{1}\left\langle \nu
_{\beta },{\bf p}_{1}\right\vert ,  \nonumber \\
\rho _{2} &=&Tr_{1}(\left\vert \psi \right\rangle \left\langle \psi
\right\vert )  \nonumber \\
&=&\left\vert C_{e}\right\vert ^{2}\left\vert \nu _{e},{\bf p}%
_{4}\right\rangle _{2}\left\langle \nu _{e},{\bf p}_{4}\right\vert
+\left\vert C_{\beta }\right\vert ^{2}\left\vert e,{\bf p}_{2}\right\rangle
_{2}\left\langle e,{\bf p}_{2}\right\vert .
\end{eqnarray}%
Under the subsequent free Hamiltonian dynamics, $\rho _{1}$ and $\rho _{2}$
evolve as%
\begin{eqnarray}
\rho _{1}^{\prime } &=&\left\vert C_{e}\right\vert ^{2}\left\vert e,{\bf p}%
_{3}\right\rangle _{1}\left\langle e,{\bf p}_{3}\right\vert +\left\vert
C_{\beta }\right\vert ^{2}\left\vert \varphi _{1},,{\bf p}_{1}\right\rangle
_{1}\left\langle \varphi _{1},{\bf p}_{1}\right\vert ,  \nonumber \\
\rho _{2}^{\prime } &=&\left\vert C_{e}\right\vert ^{2}\left\vert \varphi
_{2},{\bf p}_{4}\right\rangle _{2}\left\langle \varphi _{2},{\bf p}%
_{4}\right\vert +\left\vert C_{\beta }\right\vert ^{2}\left\vert e,{\bf p}%
_{2}\right\rangle _{2}\left\langle e,{\bf p}_{2}\right\vert ,
\end{eqnarray}%
where%
\begin{eqnarray}
\left\vert \varphi _{1}\right\rangle _{1} &=&u_{p_{1}}\left\vert \nu _{\beta
}\right\rangle _{1}+v_{p_{1}}\left\vert \nu _{e}\right\rangle _{1}, 
\nonumber \\
\left\vert \varphi _{2}\right\rangle _{2} &=&u_{p_{4}}^{\ast }\left\vert \nu
_{e}\right\rangle -v_{p_{4}}^{\ast }\left\vert \nu _{\beta }\right\rangle .
\end{eqnarray}%
$u_{p}$ and $v_{p}$ depend on time as 
\begin{eqnarray}
u_{p} &=&\cos (\lambda _{p}t)-i\frac{\Delta _{p}}{\sqrt{\lambda
_{p}^{2}+\Delta _{p}^{2}}}\sin (\lambda _{p}t),  \nonumber \\
v_{p} &=&\frac{-i\lambda _{p}}{\sqrt{\lambda _{p}^{2}+\Delta _{p}^{2}}}%
e^{i\theta _{p}}\sin (\lambda _{p}t),
\end{eqnarray}%
where $\Delta _{p}=[M_{e,e}(p)-M_{\beta ,\beta }(p)]/2$, $\lambda
_{p}=\left\vert M_{e,\beta }(p)\right\vert $, and $\theta _{p}=\arg
[M_{e,\beta }(p)]$. After this free evolution, the total $\left\vert \nu
_{e}\right\rangle $-state population is

\begin{eqnarray}
P_{\left\vert \nu _{e}\right\rangle } &=&\int d^{3}{\bf p(}_{1}\left\langle
\nu _{e},{\bf p}\right\vert \rho _{1}^{\prime }\left\vert \nu _{e},{\bf p}%
\right\rangle _{1}+_{2}\left\langle \nu _{e},{\bf p}\right\vert \rho
_{2}^{\prime }\left\vert \nu _{e},{\bf p}\right\rangle _{2})  \nonumber \\
=\left\vert C_{e}u_{p_{4}}\right\vert ^{2} &&+\left\vert C_{\beta
}v_{p_{1}}\right\vert ^{2}.
\end{eqnarray}%
Such a probability does not present the cross terms proportional to $%
C_{e}^{\ast }C_{\beta }$ and $C_{e}C_{\beta }^{\ast }$. This is due to the
fact that the state components $\left\vert \varphi _{1}\right\rangle _{1}$
and $\left\vert \varphi _{2}\right\rangle _{2}$ have different momenta and
are carried by different particles, so that quantum interference cannot
occur.

If one only concerns about the neutrino part in the two-particle system, its
behavior just after the CC reaction can be effectively described by the
classically mixed state%
\begin{equation}
\rho _{\nu }=\left\vert C_{e}\right\vert ^{2}\left\vert \nu _{e},{\bf p}%
_{4}\right\rangle \left\langle \nu _{e},{\bf p}_{4}\right\vert +\left\vert
C_{\beta }\right\vert ^{2}\left\vert \nu _{\beta },{\bf p}_{1}\right\rangle
\left\langle \nu _{\beta ,},{\bf p}_{1}\right\vert .
\end{equation}%
However, it should be born in mind that the two mixed state components are
essentially carried by two different particles. The validity of this
description can be illustrated by examining the subsequent free Hamiltonian
dynamics, by which $\rho _{\nu }$ evolves to%
\begin{equation}
\rho _{\nu }^{\prime }=\left\vert C_{e}\right\vert ^{2}\left\vert \varphi
_{2},{\bf p}_{4}\right\rangle \left\langle \varphi _{2},{\bf p}%
_{4}\right\vert +\left\vert C_{\beta }\right\vert ^{2}\left\vert \varphi
_{1},{\bf p}_{1}\right\rangle \left\langle \varphi _{1},{\bf p}%
_{1}\right\vert .
\end{equation}%
The resulting neutrino's $\left\vert \nu _{e}\right\rangle $-state
probability is also equal to $\left\vert C_{e}u_{p_{4}}\right\vert
^{2}+\left\vert C_{\beta }v_{p_{1}}\right\vert ^{2}$. This equivalence
further confirms that the CC reaction indeed destroys the quantum coherence
between $\left\vert \nu _{e}\right\rangle $ and $\left\vert \nu _{\beta
}\right\rangle $. When a second CC reaction occurs, these two particles will
be further entangled with a third particle. Under the competition between
the coherent coupling and CC-reaction-induced decoherence, the population of 
$\left\vert \nu _{e}\right\rangle $ is progressively decreased while that of 
$\left\vert \nu _{\beta }\right\rangle $ is increased until reaching the
steady state%
\begin{equation}
(\left\vert \nu _{e}\right\rangle \left\langle \nu _{e}\right\vert
+\left\vert \nu _{\beta }\right\rangle \left\langle \nu _{\beta }\right\vert
)/2.
\end{equation}%
For simplicity, we here have discarded the momentum degrees of freedom. For
this mixed state, the gain of the $\left\vert \nu _{e}\right\rangle $-state
population originating from the $\left\vert \nu _{\beta }\right\rangle
\rightarrow \left\vert \nu _{e}\right\rangle $ transition cancels out the
loss due to the $\left\vert \nu _{e}\right\rangle \longleftrightarrow
\left\vert \nu _{\beta }\right\rangle $ transition. Therefore, the
probability $P_{\left\vert \nu _{e}\right\rangle \rightarrow \left\vert \nu
_{e}\right\rangle }$ should not be smaller than 1/2, which is inconsistent
with the solar $^{8}$B neutrino experiments [1,2].

\section{FLAVOR TRANSFORMATIONS MEDIATED BY THE Z BOSONIC FIELD}

In Ref. [34], Wolfenstein proposed a neutral-current interaction that can
transform the flavor of a neutrino propagating in matter. Such an
interaction can occur only when the mediating Z bosonic field can connect
different flavor eigenstates of the neutrino. I here show that such a Z
bosonic field can lead to flavor transformations even when the neutrino
propagates in the vaccum. The dynamics of the system combined by the
neutrino field and such a Z bosonic field is governed by the Lagrangian
[28,29]

\begin{equation}
{\cal L}_{n+Z}={\cal L}_{f}+{\cal L}_{m}+{\cal L}_{I}.
\end{equation}%
${\cal L}_{f}$ is the free part without considering any coupling, which, to
the second order, can be written as%
\begin{equation}
{\cal L}_{f}=-\frac{1}{4}\left( \frac{\partial {\bf Z}_{\lambda }}{\partial
x_{\nu }}-\frac{\partial {\bf Z}_{\nu }}{\partial x_{\lambda }}\right) ^{2}-%
{\bf \phi }_{\alpha }^{\dagger }\gamma _{4}\gamma _{\lambda }\frac{\partial 
}{\partial x_{\lambda }}{\bf \phi }_{\alpha },
\end{equation}%
where ${\bf \phi }_{\alpha }$ represents the field for the neutrino with
flavor $\alpha $ ($\alpha =e,\mu ,\tau $), and ${\bf Z}_{\lambda }$ denotes
the Z bosonic field. ${\cal L}_{m}$ is the mass part, given by%
\begin{equation}
{\cal L}_{m}=-\frac{1}{2}M_{Z}^{2}{\bf Z}_{\mu }^{2},
\end{equation}%
where $M_{Z}$ is the mass of the $Z$ boson, gained by coupling to the Higgs
field. The interaction between the neutrino field and the Z bosonic field is
described by 
\begin{equation}
{\cal L}_{I}=i\eta {\bf Z}_{\lambda }U_{\alpha ,\beta }{\bf \phi }_{\alpha
}^{\dagger }\gamma _{4}\gamma _{\lambda }(1+\gamma _{5}){\bf \phi }_{\beta }.
\end{equation}%
The deviation of $U_{\alpha ,\beta }$ from $\delta _{\alpha ,\beta }$
characterizes to what extent the flavor conservation is violated due to
coupling to the Z bosonic field. The interactions associated with the W
bosonic fields are irrelevant for the neutrino flavor transformation, and
not shown here.

When $U_{\alpha ,\beta }\neq 0$, the neutrino can transform between $\alpha $%
- and $\beta $-types of flavors by emitting and then immediately
re-absorbing a virtual Z boson. To quantitatively describe thus-realized
flavor transformation, we expand ${\bf Z}_{\lambda }$ in terms of the
complete set of mode functions within a box of volume $V$ [29], 
\begin{eqnarray}
{\bf Z}_{T} &=&%
\mathop{\displaystyle\sum}%
\limits_{{\bf k}}\frac{1}{\sqrt{2\omega V}}{\bf a}_{T}{\bf (k)}e^{i{\bf %
k\cdot r}}+H.c.,  \nonumber \\
{\bf Z}_{3} &=&%
\mathop{\displaystyle\sum}%
\limits_{{\bf k}}\frac{1}{\sqrt{2\omega V}}\frac{\omega }{M_{Z}}{\bf a}_{L}%
{\bf (k)}e^{i{\bf k\cdot r}}+H.c.,  \nonumber \\
{\bf Z}_{4} &=&%
\mathop{\displaystyle\sum}%
\limits_{{\bf k}}\frac{i}{\sqrt{2\omega V}}\frac{k}{M_{Z}}{\bf a}_{L}{\bf (k)%
}e^{i{\bf k\cdot r}}+H.c.,
\end{eqnarray}%
where $\omega =\sqrt{M_{Z}^{2}+k^{2}}$, $T=1,2$ denotes the transverse
polarization degrees of freedom, and ${\bf a}_{L}({\bf k})$ and ${\bf a}_{T}(%
{\bf k})$ represent the annihilation operators for the longitudinal and
transverse polarizations with the wavevector ${\bf k}$. The neutrino field
can be expressed as%
\begin{equation}
{\bf \phi }_{\alpha }=\sqrt{\frac{1}{V}}%
\mathop{\displaystyle\sum}%
\limits_{{\bf k}}{\bf b}_{\alpha }({\bf k})e^{i{\bf k\cdot r}},
\end{equation}%
where ${\bf b}_{\alpha }$ denote the annihilation operator of the neutrino
field with the flavor $\alpha $. As we here focus on the left-handed
neutrino, the right-handed anti-neutrino part is not included in ${\bf \phi }%
_{\alpha }$. Then the Hamiltonian for neutrino and the Z bosonic field is
given by%
\begin{equation}
{\cal H}_{n+Z}={\cal H}_{0}+{\cal H}_{I},
\end{equation}%
Here ${\cal H}_{0}$ corresponds to the "free" part without considering the
neutrino-boson interaction, given by [29] 
\begin{equation}
{\cal H}_{0}=%
\mathop{\displaystyle\sum}%
\limits_{{\bf k}}k{\bf b}_{\alpha }^{\dagger }({\bf k}){\bf b}_{\alpha }(%
{\bf k})+%
\mathop{\displaystyle\sum}%
\limits_{{\bf k}}\omega \left[ {\bf a}_{L}^{\dagger }{\bf (k)a}_{L}{\bf (k)}+%
\mathop{\displaystyle\sum}%
\limits_{T=1,2}{\bf a}_{T}^{\dagger }{\bf (k)a}_{T}{\bf (k)}\right] .
\end{equation}%
The interaction part ${\cal H}_{I}$ is 
\begin{eqnarray}
{\cal H}_{I} &=&-\int dx{\cal L}_{I}  \nonumber \\
&=&-\frac{i\eta }{\sqrt{2\omega V}}U_{\alpha ,\beta }%
\mathop{\displaystyle\sum}%
\limits_{{\bf k,k}^{\prime }}{\bf X}_{\lambda }{\bf b}_{\alpha }^{\dagger }(%
{\bf k})\gamma _{4}\gamma _{\lambda }(1+\gamma _{5}){\bf b}_{\beta }({\bf k}%
^{\prime }),
\end{eqnarray}%
where%
\begin{eqnarray}
{\bf X}_{T} &=&{\bf a}_{T}{\bf (k-k^{\prime })}+{\bf a}_{T}^{\dagger }{\bf (k%
}^{\prime }{\bf -k)},  \nonumber \\
{\bf X}_{3} &=&\frac{\omega ^{\prime }}{M_{Z}}\left[ {\bf a}_{L}{\bf %
(k-k^{\prime })}+\omega ^{\prime }{\bf a}_{T}^{\dagger }{\bf (k}^{\prime }%
{\bf -k)}\right] ,  \nonumber \\
{\bf X}_{4} &=&\frac{i\left\vert {\bf k-k^{\prime }}\right\vert }{M_{Z}}%
\left[ {\bf a}_{L}{\bf (k-k^{\prime })}-{\bf a}_{T}^{\dagger }{\bf (k}%
^{\prime }{\bf -k)}\right] ,
\end{eqnarray}%
with $\omega ^{\prime }=\sqrt{M_{Z}^{2}+\left\vert {\bf k-k^{\prime }}%
\right\vert ^{2}}$.

As the energy of the neutrino is much smaller than $M_{Z}$, the neutrino
cannot emit any real Z gauge boson. However, the virtual excitation of the Z
boson can induce effective couplings among different neutrino flavors.
Suppose that the system is initially in the state $\left\vert \Psi
_{i}\right\rangle =\left\vert \nu _{\alpha }({\bf p})\right\rangle
\left\vert 0\right\rangle $, where ${\bf p}$ denotes the momentum of the
neutrino, and $\left\vert 0\right\rangle $ is the vacuum state of the Z
boson field. This initial state can be coupled to the final state $%
\left\vert \Psi _{f}\right\rangle =\left\vert \nu _{\beta }({\bf p}%
),0\right\rangle $ through the intermediate states $\left\vert \Psi _{m}^{j}(%
{\bf k})\right\rangle ={\bf a}_{j}^{\dagger }({\bf p}-{\bf k})\left\vert \nu
_{\gamma }({\bf k}),0\right\rangle $, where $j=T,L$. Based on the
second-order perturbation theory, ${\cal H}_{int}$ can be replaced by the
effective Hamiltonian%
\begin{equation}
{\cal H}_{eff}=\xi _{\alpha ,\beta }\left\vert \nu _{\alpha }({\bf p}%
),0\right\rangle \left\langle \nu _{\beta }({\bf p}),0\right\vert ,
\end{equation}%
where 
\begin{eqnarray}
\xi _{\alpha ,\beta } &\simeq &-%
\mathop{\displaystyle\sum}%
\limits_{{\bf k}}\frac{\left\langle \Psi _{i}\right\vert {\cal H}%
_{int}\left\vert \Psi _{m}^{j}({\bf k})\right\rangle \left\langle \Psi
_{m}^{j}({\bf k})\right\vert {\cal H}_{int}\left\vert \Psi _{f}\right\rangle 
}{M_{z}}  \nonumber \\
&\simeq &-%
\mathop{\displaystyle\sum}%
\limits_{{\bf k}}\frac{6\eta ^{2}U_{\alpha ,\gamma }U_{\gamma ,\beta }}{%
M_{z}^{2}V}  \nonumber \\
&=&-\frac{3\eta ^{2}}{4\pi ^{3}M_{z}^{2}}\int d^{3}{\bf k}U_{\alpha ,\gamma
}U_{\gamma ,\beta }.
\end{eqnarray}%
The integral divergence can be eliminated if $U_{\alpha ,\gamma }$ contains
a decaying factor, such as $e^{-\Gamma \left\vert {\bf k-p}\right\vert ^{2}}$%
, which implies that the coupling between the initial and intermediate
neutrino states exponentially decreases with the increasing momentum-phase
distance. The result shows that the neutrino can change its flavor through
virtual excitation of Z bosons, which can be understood as a backaction of
the gauge field on the neutrino.

\section{INTERPRETATION OF EXPERIMENTAL RESULTS BASED ON THE NEW MECHANISM}

We now proceed to qualitatively interpret previous neutrino experiments
based on this physical mechanism. Discarding the degrees of freedom of the Z
bosonic field and performing the transformation $e^{i(p+\xi _{\tau ,\tau
})t} $, the effective Hamiltonian ${\cal H}_{eff}$ is transformed to 
\begin{eqnarray}
{\cal H}_{eff}^{\prime } &=&\Delta \left\vert \nu _{e}\right\rangle
\left\langle \nu _{e}\right\vert +(\Delta +\delta )\left\vert \nu _{\mu
}\right\rangle \left\langle \nu _{\mu }\right\vert  \nonumber \\
&&+(\xi _{e,\mu }\left\vert \nu _{e}\right\rangle \left\langle \nu _{\mu
}\right\vert +\xi _{\tau ,\mu }\left\vert \nu _{\tau }\right\rangle
\left\langle \nu _{\mu }\right\vert  \nonumber \\
&&+\xi _{\tau ,e}\left\vert \nu _{\tau }\right\rangle \left\langle \nu
_{e}\right\vert +H.c.),
\end{eqnarray}%
where $\Delta =\xi _{e,e}-\xi _{\tau ,\tau }$ and $\delta =\xi _{\mu ,\mu
}-\xi _{e,e}$. We here suppose all the Hamiltonian parameters are real.
Under the condition $\delta ,\xi _{e,\mu }\ll \xi _{\tau ,\alpha }\ll \Delta 
$ with $\alpha =e,\mu $, the neutrino evolution exhibits both fast and slow
oscillations. For the short timescale $T\ll 1/\lambda $ with $\lambda =\xi
_{e,\mu }+\xi _{\tau ,e}\xi _{\tau ,\mu }/\Delta $, the transition
probability from the initial state $\left\vert \nu _{\alpha }\right\rangle $
($\alpha =e,\mu $) to $\left\vert \nu _{\tau }\right\rangle $, caused by the
fast oscillation, is approximately given by 
\begin{equation}
P_{f}(\left\vert \nu _{\alpha }\right\rangle \rightarrow \left\vert \nu
_{\tau }\right\rangle )\simeq \frac{1}{2}A_{f}^{\alpha }[1-\cos (\Delta t)],
\end{equation}%
where $A_{f}^{\alpha }=4\xi _{\tau ,\alpha }^{2}/\Delta ^{2}$. When $%
A_{s}^{\alpha }\ll 1$, such an oscillation has a small amplitude. The
electron-antineutrino disappearance observed in the reactor neutrino
experiments [17-19] can be described by this equation. The slow oscillation
occur between $\left\vert \nu _{e}\right\rangle $ and $\left\vert \nu _{\mu
}\right\rangle $. The resulting transition probability is well approximated
by 
\begin{equation}
P_{s}(\left\vert \nu _{e}\right\rangle \longleftrightarrow \left\vert \nu
_{\mu }\right\rangle )\simeq \frac{1}{2}A_{s}[1-\cos (\Omega _{l}t)].
\end{equation}%
where $\Omega =\sqrt{4\lambda ^{2}+\delta ^{\prime 2}}$ and $%
A_{s}=\left\vert 2\lambda /\Omega \right\vert ^{2}$ with $\delta ^{\prime
}=\delta +(\xi _{\tau ,\mu }^{2}-\xi _{\tau ,e}^{2})/\Delta $. For $\lambda
\gtrsim \left\vert \delta ^{\prime }\right\vert $, we have $A_{s}\sim 1$,
which qualitatively coincides with the KamLAND experimental result [20,21],
although the quantitative dependences of the Hamiltonian parameters $\xi
_{\alpha ,\beta }$ on the neutrino energy are still unclear.

The observed deficit of solar electron neutrinos can be well explained in
terms of the competition between the coherent couplings induced by the Z
bosonic field and the decoherence effect due to the CC interaction, which
can be modeled as the incoherent transformation 
\begin{equation}
\rho \rightarrow S_{e}\rho S_{e}+S_{\mu +\nu }\rho S_{\mu +\nu },
\end{equation}%
where $S_{e}=\left\vert \nu _{e}\right\rangle \left\langle \nu
_{e}\right\vert $, $S_{\mu +\nu }=\left\vert \nu _{\mu }\right\rangle
\left\langle \nu _{\mu }\right\vert +\left\vert \nu _{\tau }\right\rangle
\left\langle \nu _{\tau }\right\vert $, and $\rho $ denotes the density
operator of the neutrino before the CC reaction. Phenomenally, the
competition between the coherent flavor coupling and incoherent scattering
associated with $\left\vert \nu _{e}\right\rangle $ can be described by the
master equation%
\begin{equation}
\frac{d\rho }{dt}=i\left[ \rho ,{\cal H}_{eff}^{\prime }\right] +\frac{%
\gamma }{2}(2S_{e}\rho S_{e}-S_{e}\rho -S\sigma _{e}).
\end{equation}%
The decoherence rate $\gamma $ depends upon the neutrino energy and the
properties of the matter. As the scattering cross-section scales with the
neutrino energy, the high-energy neutrinos would have evolved into a steady
state before reaching the solar surface, described by the density operator $%
\rho =\frac{1}{3}%
\mathop{\displaystyle\sum}%
\limits_{\alpha }\left\vert \nu _{\alpha }\right\rangle \left\langle \nu
_{\alpha }\right\vert $, which remains invariant during the subsequent
propagation. Consequently, the proportion of detected $\left\vert \nu
_{e}\right\rangle \rightarrow \left\vert \nu _{e}\right\rangle $ events is $%
1/3$, as revealed in solar $^{8}$B neutrino experiments [1,2]. If the
neutrino is free of such incoherent scatterings, the average fraction of the 
$\left\vert \nu _{e}\right\rangle $-outcomes after a long-time propagation
is $1-\frac{1}{2}A_{s}$. For the intermediate case, the value is between
these two extremes [1].

The atmospheric neutrinos originate from the pion decay and subsequent muon
decay, which produce both the $\mu $- and $e$-type neutrinos with the
numbers $N_{\mu }^{0}\simeq 2N_{e}^{0}$. According to the Super-Kamiokande
experiments [10,11], when the Zenith angle ($\Theta $) ranges from $0^{\circ
}$ to $80^{\circ }$, the disappearance probability of the $\mu $-type events
for multi-GeV neutrinos exhibits slight oscillations, reflecting the
comprehensive effect of the fast oscillation to $\left\vert \nu _{\tau
}\right\rangle $ and the slow $\left\vert \nu _{e}\right\rangle
\longleftrightarrow \left\vert \nu _{\mu }\right\rangle $ oscillation. Due
to the imbalance between $N_{\mu }^{0}$ and $N_{e}^{0}$, for the number of $%
\mu $-type events ($N_{\mu }$) the gain from the $\left\vert \nu
_{e}\right\rangle \rightarrow \left\vert \nu _{\mu }\right\rangle $
transition is less than the loss due to $\left\vert \nu _{\mu }\right\rangle
\rightarrow \left\vert \nu _{e}\right\rangle $ transition. For $N_{e}$, the
loss due to $\left\vert \nu _{e}\right\rangle \rightarrow \left\vert \nu
_{\tau }\right\rangle $ transfer is partly compensated for by the net gain
from the $\left\vert \nu _{e}\right\rangle \longleftrightarrow \left\vert
\nu _{\mu }\right\rangle $ oscillation, so that its deficit is much smaller
than that of $N_{\mu }$. With the extension of the propagation distance in
the vacuum, the slow $\left\vert \nu _{e}\right\rangle \longleftrightarrow
\left\vert \nu _{\mu }\right\rangle $ oscillation plays an increasingly
important role. Within the regime $80^{\circ }<\Theta <90^{\circ }$, the
number of net $\left\vert \nu _{\mu }\right\rangle \rightarrow \left\vert
\nu _{e}\right\rangle $ transfer events surpasses that of $\left\vert \nu
_{e}\right\rangle \rightarrow \left\vert \nu _{\tau }\right\rangle $ ones.
When $\Theta >90^{\circ }$, the neutrino evolves under the competition
between the coherent Hamiltonian dynamics and decoherence effect due to
interactions with matter of the Earth. For multi-GeV neutrinos, the
reactions $\nu _{\mu /\tau }+e\rightarrow \mu /\tau +\nu _{e}$ are allowed,
but their occurrence probabilities would be much smaller than that for $\nu
_{e}+e\rightarrow e+\nu _{e}$, since the transformations from $\nu _{\mu
/\tau }$ into $\mu $ need to cost much more kinetic energy than the $\nu
_{e}\rightarrow e$ transformation. As the sensitivities of the charged- and
neutral-current reactions strongly depend upon the type of the neutrino,
they will deteriorate the coherence between the neutrino flavors. For $%
90^{\circ }<\Theta <100^{\circ }$, the neutrino retains some quantum
coherence after passing through the Earth, exhibiting slight oscillations.
With the increase of $\Theta $, the slow $\left\vert \nu _{e}\right\rangle
\longleftrightarrow \left\vert \nu _{\mu }\right\rangle $ oscillation has a
higher probability of being interrupted by scattering events, which prevents
monotonous increase of the net gain of $N_{e}$ from such an oscillation.
When $100^{\circ }<\Theta <180^{\circ }$, the neutrino has approximately
evolved to the maximally mixed state $\rho $ after crossing the Earth, so
that the three flavors roughly have the same population, independent of the
initial state.

\section{DISCUSSIONS}

As our theoretical model allows occurrence of oscillations even if the
neutrino is massless, one may argue that the state of a massless particle
cannot be changed during its propagation in the vacuum. This is the case if
the particle remains unperturbed, as exemplified by the free propagation of
a photon. However, the photonic state can be changed by some scattering
process. For example, a photon can transform between modes $a$ and $b$ with
the assistance of $\Lambda $-type three-level atoms, which have an excited
state $\left\vert e\right\rangle $ and two ground states $\left\vert
g\right\rangle $ and $\left\vert f\right\rangle $ [46]. The transitions $%
\left\vert e\right\rangle \rightarrow \left\vert g\right\rangle $ and $%
\left\vert e\right\rangle \rightarrow \left\vert f\right\rangle $ are
respectively coupled to these two modes with the same detuning. When the
detuning is much larger than the coupling strengths, the photon would
oscillate between these two modes through a Raman scattering process,
mediated by virtual excitation of the atoms. The flavor oscillations of the
neutrino bear some similarity to such a Raman process. As the neutrino is
inevitably disturbed by the vacuum fluctuations of the gauge fields
pervading the space, its propagation cannot be really free even in the
vacuum. The backaction of the virtually excited Z gauge field on the
neutrino is analogous to that of the virtually excited atoms on the photon
in the Raman process. One may further ask: Does there exist any other
mechanism that can mediate the neutrino flavor transformation? The answer is
presumably negative for the following reason. The W and Z bosonic fields are
the only elements that can interact with the neutrino during its propagation
in the vacuum. As the neutrino flavor is defined by the charged current that
is mediated by the W bosonic field, the vacuum fluctuation of the Z bosonic
field seems to be the only mechanism that can change the neutrino flavor.
Finally, it should be noted that the virtual excitation of Z bosons cannot
lead to flavor transformations between different flavors of charged leptons
owing to their large mass differences.

\end{document}